\documentclass[prb,twocolumn,showkeys,showpacs,amsmath,amsfonts,floatfix,superscriptaddress,nofootinbib]{revtex4-1}

\usepackage{amssymb,amsmath,amstext}                
\usepackage{graphicx}                                               
\usepackage{epstopdf}                                               
\usepackage{color}       
\usepackage{bm}
\usepackage{appendix}                                              
\usepackage[latin2]{inputenc}
\usepackage{ulem}
\usepackage{latexsym}
\usepackage[colorlinks=true,citecolor=blue,linkcolor=magenta]{hyperref}
\def\be{\begin{equation}}
\def\ee{\end{equation}}
\def\bea{\begin{eqnarray}}
\def\eea{\end{eqnarray}}
\def\bi{\begin{itemize}}
\def\ei{\end{itemize}}

\newcommand{\bra}[1]{\mbox{$\langle #1 |$}}
\newcommand{\ket}[1]{\mbox{$| #1 \rangle$}}

\begin{document}

\title{ Tensor network simulation of the quantum Kibble-Zurek quench\\ 
        from the Mott to superfluid phase in the two-dimensional Bose-Hubbard model }

\author{Jacek Dziarmaga} 
\affiliation{Jagiellonian University, 
             Faculty of Physics, Astronomy and Applied Computer Science,
             Institute of Theoretical Physics, 
             ul. \L{}ojasiewicza 11, 30-348 Krak\'ow, Poland }  
\affiliation{Jagiellonian University, 
             Mark Kac Center for Complex Systems Research,
             ul. \L{}ojasiewicza 11, 30-348 Krak\'ow, Poland }  

\author{Jakub M. Mazur} 
\affiliation{Jagiellonian University, 
             Faculty of Physics, Astronomy and Applied Computer Science,
             Institute of Theoretical Physics, 
             ul. \L{}ojasiewicza 11, 30-348 Krak\'ow, Poland }               

\date{February 26, 2023}

\begin{abstract}
Quantum simulations of the Bose-Hubbard model (BHM) at commensurate filling can follow spreading of correlations after a sudden quench for times long enough to estimate their propagation velocities.
In this work we perform tensor network simulation of the quantum Kibble-Zurek (KZ) ramp from the Mott towards the superfluid phase in the square lattice BHM and demonstrate that even relatively short ramp/quench times allow one to test the power laws predicted by the KZ mechanism (KZM). They can be verified for the correlation length and the excitation energy but the most reliable test is based on the KZM scaling hypothesis for the single particle correlation function: scaled correlation functions for different quench times evaluated at the same scaled time collapse to the same scaling function of the scaled distance. The scaling of the space and time variables is done according to the KZ power laws. 
\end{abstract}

\maketitle

\section{Quantum Kibble-Zurek mechanism}
\label{sec:intro}


The Kibble-Zurek mechanism (KZM) originated from a scenario for topological defect formation in cosmological phase transitions driven by expanding and cooling Universe~\cite{K-a, *K-b, *K-c}. Kibble considered independent selection of broken symmetry vacua in causally disconnected regions. The result is a mosaic of broken symmetry domains, whose size is limited by the causal horizon, leading to topologically nontrivial configurations. 
However, the speed of light is not relevant for laboratory experiments in condensed matter systems where, instead, a dynamical theory for the continuous phase transitions~\cite{Z-a,*Z-b,*Z-c,Z-d} predicts the scaling of the defects density as a function of the quench rate employing equilibrium critical exponents. It has been verified by numerous simulations~\cite{KZnum-a,KZnum-b,KZnum-c,KZnum-d,KZnum-e,KZnum-f,KZnum-g,*KZnum-h,*KZnum-i,KZnum-j,KZnum-k,KZnum-l,KZnum-m,that} and condensed matter experiments~\cite{KZexp-a,KZexp-b,KZexp-c,KZexp-d,KZexp-e,KZexp-f,KZexp-g,KZexp-gg,KZexp-h,KZexp-i,KZexp-j,KZexp-k,KZexp-l,KZexp-m,KZexp-n,KZexp-o,KZexp-p,KZexp-q,KZexp-r,KZexp-s,KZexp-t,KZexp-u,KZexp-v,KZexp-w,KZexp-x}. Topological defects play central role in these studies as they survive inevitable dissipation.

Their role was played down in the quantum KZM (QKZM) that considers quenches across quantum critical points in isolated quantum systems~\cite{QKZ1,QKZ2,QKZ3,d2005,d2010-a, d2010-b, QKZteor-a,QKZteor-b,QKZteor-c,QKZteor-d,QKZteor-e,QKZteor-f,QKZteor-g,QKZteor-h,QKZteor-i,QKZteor-j,QKZteor-k,QKZteor-l,QKZteor-m,QKZteor-n,QKZteor-o,KZLR1,KZLR2,QKZteor-oo,delcampostatistics,KZLR3,QKZteor-q,QKZteor-r,QKZteor-s,QKZteor-t,sonic,QKZteor-u,QKZteor-v,QKZteor-w,QKZteor-x, roychowdhury2020dynamics,sonic,schmitt2021quantum,RadekNowak,dziarmaga_kinks_2022,oscillations}. It was tested by experiments~\cite{QKZexp-a, QKZexp-b, QKZexp-c, QKZexp-d, QKZexp-e, QKZexp-f, QKZexp-g,deMarco2,Lukin18,adolfodwave,2dkzdwave,King_Dwave1d_2022,Semeghini2021,Satzinger2021etal}. Recent developments in Rydberg atoms' quantum simulators ~\cite{rydberg2d1,rydberg2d2,Semeghini2021,Satzinger2021etal} and coherent D-Wave~\cite{King_Dwave1d_2022,King_Dwave_glass} open possibility to study the QKZM in two and three spatial dimensions and/or to employ it as a test of quantumness of the simulator~\cite{RadekNowak, King_Dwave1d_2022, dziarmaga_kinks_2022, schmitt2021quantum,
oscillations}.


The QKZM can be described in brief as follows. 
A smooth ramp crossing the critical point at time $t=0$ can be linearized in its vicinity as 
\begin{equation}
\epsilon(t)=\frac{t}{\tau_Q}.
\label{epsilont}
\end{equation}
Here $\epsilon$ is a dimensionless parameter in a Hamiltonian, that measures distance from the quantum critical point, and $\tau_Q$ is called a quench time. Initially, the system is prepared in its ground state far from the critical point. At first the evolution adiabatically follows the ground state of the changing Hamiltonian until the adiabaticity fails near time $-\hat t$ when the energy gap becomes comparable to the ramp rate: $\Delta\propto|\epsilon|^{z\nu} \propto |\dot \epsilon/\epsilon| = 1/|t|$. This KZM timescale is 
\be 
\hat t\propto \tau_Q^{z\nu/(1+z\nu)}.
\label{hatt}
\ee
Here $z$ and $\nu$ are the dynamical and the correlation length critical exponents, respectively. 

From a causality point of view~\cite{Z-a, *Z-b, *Z-c,sonic}, which is most straightforward when the dynamical exponent $z=1$ and the excitations have a definite speed of sound at the critical point, the correlation length initially grows as $\xi\propto|\epsilon|^{-\nu}$ in step with the correlation length in the adiabatic ground state, that would eventually diverge at the critical point, but near $-\hat t$ its diverging growth rate, 
\be
    \frac{d\xi}{dt}=
    \frac{d\epsilon}{dt}
    \frac{d\xi}{d\epsilon}\propto
    \tau_Q^{-1}
    \frac{1}{|\epsilon|^{\nu+1}},
\ee
exceeds the speed limit at which correlations can spread near the critical point. The following growth is limited by $2c$~\cite{sonic}, where $c$ is the relevant speed of sound at the critical point. The correlation length at $-\hat t$, 
\begin{equation}
\hat\xi \propto \tau_Q^{\nu/(1+z\nu)}, 
\label{hatxi}
\end{equation}
defines the characteristic KZ length. Despite the following growth between $-\hat t$ and $0$, the correlation range when crossing the critical point is also proportional to $\hat\xi$ although usually a few times longer~\cite{sonic}. The causality picture can be generalized to $z\neq1$ where $c$ has to be replaced by a relevant speed of excitations that depends on $\tau_Q$~\cite{sonic}.

The two KZ scales are interrelated by
\be   
    \hat t \propto \hat\xi^z.
\ee 
Accordingly, in the KZM regime after $-\hat t$, observables are expected to satisfy the KZM dynamical scaling hypothesis~\cite{KZscaling1,KZscaling2,Francuzetal} with $\hat\xi$ being the unique scale. For, say, a two-point observable ${\cal O}_r$, where $r$ is a distance between the two points, it reads
\be 
\hat\xi^{\Delta_{\cal O}} \bra{\psi(t)} {\cal O}_r \ket{\psi(t)} = 
F_{\cal O}\left( t/\hat\xi^z , R/\hat\xi \right),
\label{KZscalingO}
\ee
where $\ket{\psi(t)}$ is the state during the quench, $\Delta_{\cal O}$ is the scaling dimension, and $F_{\cal O}$ is a non-universal scaling function. 

In this paper we consider the QKZM in the 2D Bose-Hubbard model (BHM) on an infinite square lattice. We assume commensurate filling of one particle per site with a well defined Mott-superfluid quantum phase transition. A sudden quench from deep in the Mott phase to the superfluid side of the transition was studied both experimentally~\cite{exp_sudden_BH_2D} and numerically~\cite{BH2Dcorrelationspreading}. After the quench the system was allowed to evolve with the final Hamiltonian for a time long enough to estimate the speed at which correlations were spreading -- the central phenomenon in the causal interpretation of the QKZM. The aim of the present paper is to demonstrate numerically that these evolution times would also be long enough to verify the KZM scaling hypothesis.  

The experimental set up \cite{exp_sudden_BH_2D}, where the initial state is a Mott state with the commensurate $n=1$ particle per site, provides an opportunity to go beyond the previous experimental test \cite{QKZexp-e} where the initial atomic cloud had non-uniform occupation numbers in the range $n=1..3$. $n\approx 3$ in the center of the trap may be just large enough to explain why the measured power laws for relatively fast quenches were consistent with the QKZM but with the mean-field values of the critical exponents.   
Another attempt was made in Ref. \onlinecite{Braun_2015} but a limited range of available parameters made the experimental results inconclusive, though in good agreement with numerical simulations of the experimental set-up.
On the numerical front a more tractable 1D version was considered~\cite{KZM1DBHM,KZM1DBHM_Gardas} where the Kosterlitz-Thouless nature of the transition makes $\hat\xi$ only logarithmically dependent on $\tau_Q$ and, therefore, a clear-cut test of the KZM would require quench times ranging over many orders of magnitude. In contrast, the 2D transition is sharper, the KZM power laws are steeper and their experimental verification should be unambiguous. However, numerical simulation of the non-integrable 2D model is more demanding as the applicability of the numerically exact tensor-network DMRG-like methods becomes severely limited in 2D and one may be forced to resort to the mean-field Gutzwiller ansatz\cite{BH_Gutzwiller} instead.
In this work we overcome the limitations of the quasi-1D DMRG by employing a genuine 2D tensor network.  

\section{2D tensor network algorithm}
\label{sec:ntu}

Typical quantum many body states can be represented efficiently by tensor networks~\cite{Verstraete_review_08,Orus_review_14}. These include the matrix product states (MPS) in one dimension (1D)~\cite{fannes1992}, the projected entangled pair state (PEPS) in 2D~\cite{Nishino_2DvarTN_04,verstraete2004}, or the multi-scale entanglement renormalization ansatz (MERA)~\cite{Vidal_MERA_07,Vidal_MERA_08,Evenbly_branchMERA_14,Evenbly_branchMERAarea_14} incarnating the real space renormalization group.
Recently an infinite PEPS ansatz (iPEPS) was employed to simulate unitary time evolution on infinite lattices\cite{CzarnikDziarmagaCorboz,HubigCirac,tJholeHubig,Abendschein08,SUlocalization,SUtimecrystal,ntu,mbl_ntu,schmitt2021quantum,BH2Dcorrelationspreading,ising2D_correlationsperading}. The simulations include spreading of correlations after a sudden quench in the Bose-Hubbard model (BHM)\cite{BH2Dcorrelationspreading} and the transverse field Ising model \cite{ising2D_correlationsperading} as well as the KZ ramp in the latter \cite{schmitt2021quantum}. In this work we perform simulations of the KZ ramp in the BHM that seem timely in view of the new opportunities opened by the recent experiment ~\cite{exp_sudden_BH_2D}.  

\begin{figure}[t!]
\vspace{-0cm}
\includegraphics[width=0.999\columnwidth,clip=true]{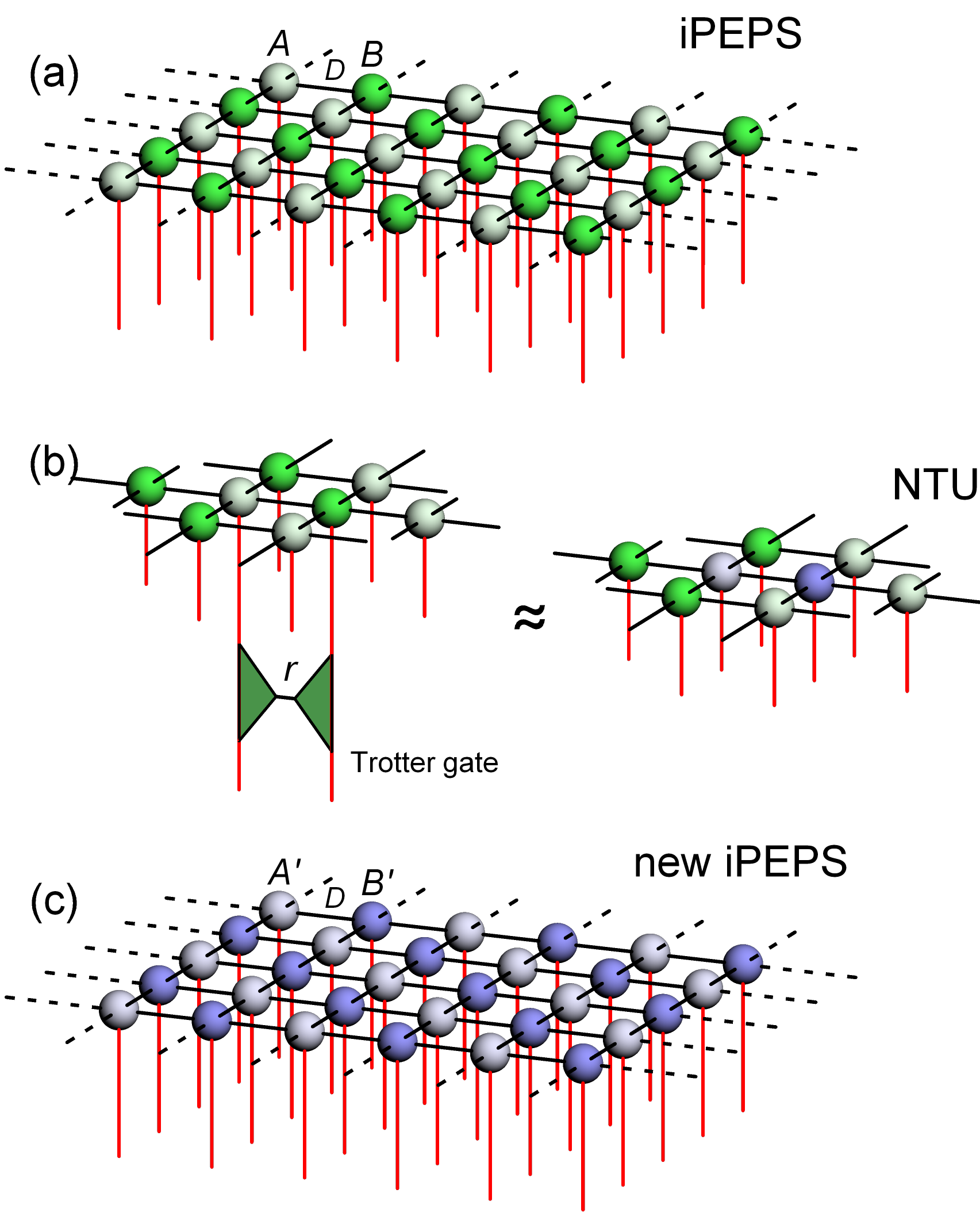}
\vspace{-0cm}
\caption{
{\bf Essential NTU. }
In (a) infinite PEPS with tensors $A$ (lighter green) and $B$ (darker green) on the two sublattices.
The red lines are physical spin indices and the black lines are bond indices, with bond dimension D, 
contracting NN sites. 
In one of Suzuki-Trotter steps a Trotter gate is applied to every NN pair of $A$-$B$ tensors along every horizontal row (but not to horizontal $B$-$A$ pairs). The gate can be represented by a contraction of two tensors by an index with dimension $r$. When the two tensors are absorbed into tensors $A$ and $B$ the bond dimension between them increases from $D$ to $r\times D$.
In (b) the $A$-$B$ pair -- with a Trotter gate applied to it -- is approximated by a pair of new tensors, $A'$ (lighter blue) and $B'$ (darker blue), connected by an index with the original dimension $D$. The new tensors are optimized to minimize difference between the two networks in (b).
After $A'$ and $B'$ are converged, they replace all tensors $A$ and $B$ in a new iPEPS shown in (c). 
Now the next Trotter gate can be applied.
The dominant numerical cost of the NTU procedure scales as $D^8$ and is fully parallelizable \cite{ntu}.
}
\label{fig:NTU}
\end{figure}

We apply the neighbourhood tensor update (NTU) algorithm \cite{ntu} that was previously used to simulate the many body localization \cite{mbl_ntu} and the KZ ramp in the Ising model \cite{schmitt2021quantum}. The evolution operator is Suzuki-Trotter decomposed \cite{Trotter_59,Suzuki_66,Suzuki_76} into a product of nearest neighbor(NN) Trotter gates. As each Trotter gate increases the bond dimension along its NN bond, it has to be truncated back to its original value to prevent its exponential growth with time. The truncation has to be done in a way that minimizes an error afflicted to the quantum state. There are several numerical error measures, each of them implying a different algorithm: the simple update (SU) \cite{tJholeHubig,SUlocalization}, the full update (FU) \cite{fu,CzarnikDziarmagaCorboz}, the neighbourhood tensor update (NTU) \cite{ntu,schmitt2021quantum,mbl_ntu}, or gradient tensor update (GTU) \cite{gradient}. The NTU error measure is explained in Fig. \ref{fig:NTU}. This is the efficient and stable algorithm to be employed here.

In each Trotter gate the Frobenius norm of the difference between the left ($L$) and right ($R$) hand sides of Fig. \ref{fig:NTU}b is minimized. The norm,
\be 
\delta = || L-R ||,
\label{eq:delta}
\ee 
is what we call an NTU error. For small enough time step it should become proportional to $dt$. $\delta$ is an estimate for an error inflicted on local observables by the bond dimension truncation. Accumulating Trotter errors can eventually derail the time evolution. In the worst case scenario the errors are additive. This motivates an integrated NTU error \cite{Hubbard_Sinha}, 
\be 
\Delta = \sum_i \delta_i,
\ee 
where the sum is over all performed Trotter gates. For a second order Suzuki-Trotter decomposition on a bipartite square lattice, where each time step is a sequence of $8$ NN Trotter gates, which is 4 gates per site, $4\Delta$ estimates an error of a typical local observable. The observables are calculated with the help of the corner transfer matrix renormalization group \cite{corboz14_tJ,corboz16b}.

\section{Bose Hubbard model}
\label{sec:bh}

The Hamiltonian on an infinite square lattice is
\bea
H &=& 
- J\sum_{\langle i,j \rangle} \left( b_{i}^\dag b_{j} + b_{j}^\dag b_{i} \right) + 
  \frac{U}{2} \sum_i  n_i\left( n_{i} - 1 \right).
\label{H}
\eea
Here $b_{i}^\dag$ and $b_{i}$, respectively, creates and annihilates a boson on site $i$, $n_{i}=b^\dag_{i}b_{i}$ is the number operator, $J$ is the strength of the hopping between nearest-neighbor sites, and $U$ is on-site repulsion strength. $\langle i,j \rangle$ denotes summation over nearest-neighbor (NN) pairs in the hopping energy (every pair contributes to the sum only once). For the commensurate filling of $n=1$ particles per site the continuous Mott-superfluid quantum phase transition is located at $U/J=16.7$~\cite{ElstnerMonien,BHM_Svistunov,Krutitsky_review}. The dynamical exponent $z=1$ and the correlation length exponent $\nu=0.67$, hence $\hat\xi\propto\tau_Q^{0.40}$. 

In an optical lattice both $J$ and $U$ depend on the recoil energy. Deep in the tight binding regime the dependence of $J$ is roughly exponential while that of $U$ is relatively weak (if not negligible). In a tensor network simulation the dimension of the local Hilbert space has to be truncated to a finite physical dimension $d$, i.e., to occupation numbers $0,...,d-1$. This is self-consistent on the Mott side of the transition, including the critical point, thanks to limited variance of occupation numbers $n_i$.

\begin{figure}[t!]
\vspace{-0cm}
\includegraphics[width=0.999\columnwidth,clip=true]{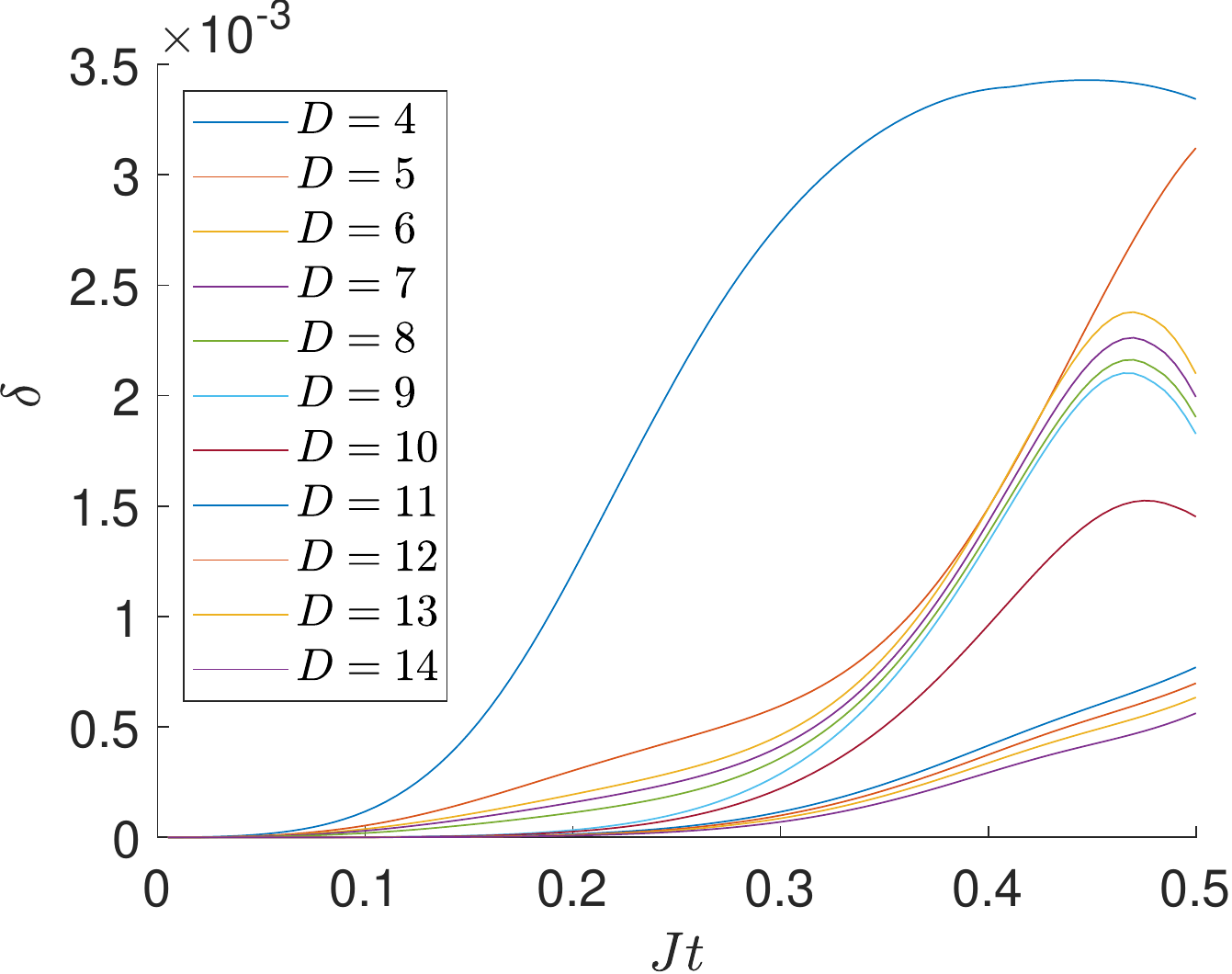}
\includegraphics[width=0.999\columnwidth,clip=true]{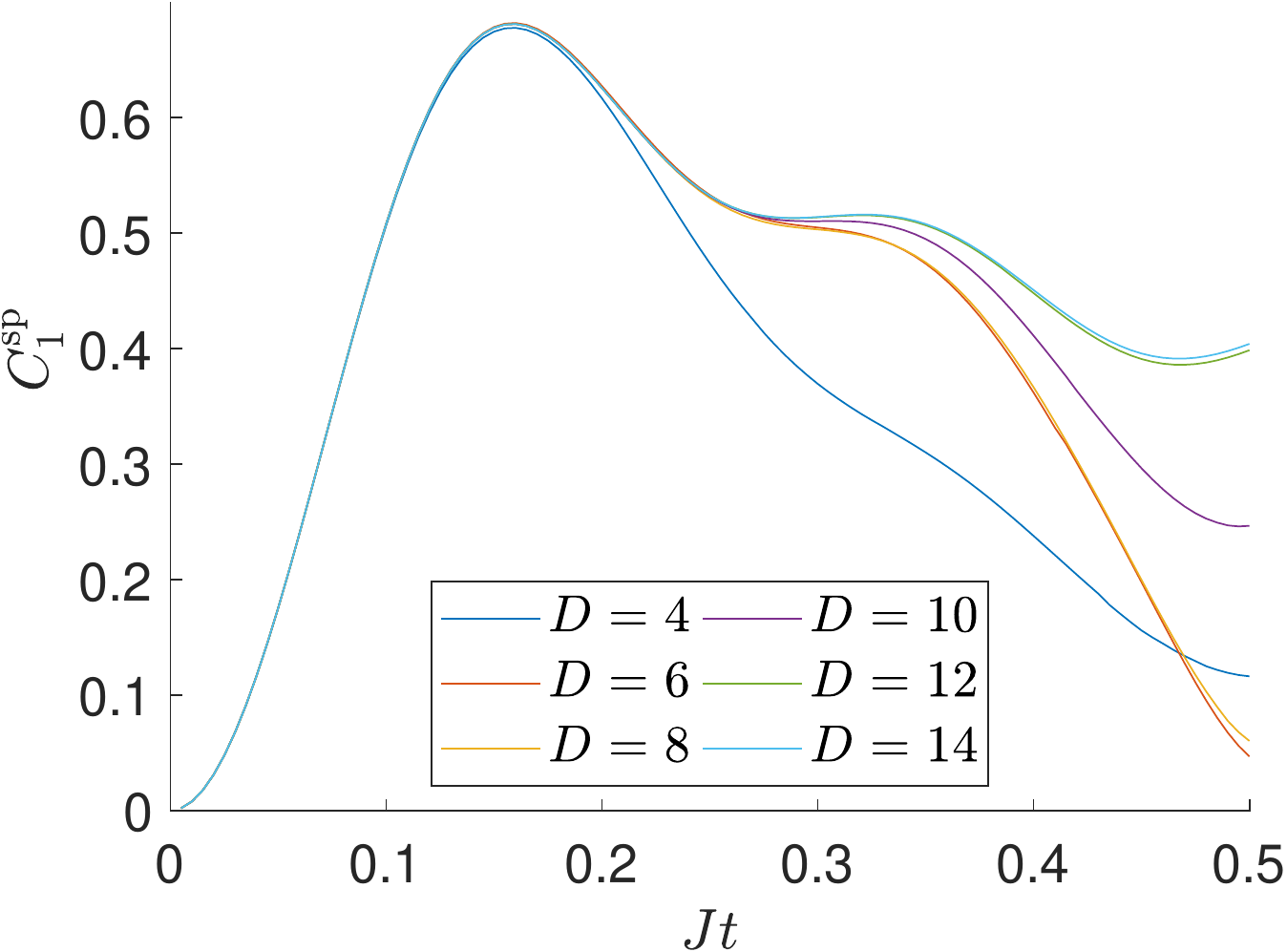}
\vspace{-0cm}
\caption{
{\bf Sudden quench to $U/J=19.6$. }
In (a) the NTU error and in (b) the NN single particle correlator in function of time.
The correlator appears converged in $D$ already for $D=6..8$ but the NTU error in this range is still unacceptable ($4\Delta\approx 0.1$) and, indeed, for higher $D=11..14$ the correlator finds a new converged curve, this time with acceptable errors ($4\Delta\approx 0.01$).
Here we set $J=1$, $U=19.6$, $Jdt=0.005$, and physical dimension $d=3$.
}
\label{fig:sudden}
\end{figure}

\section{Sudden quench revisited}
\label{sec:quench}

As a benchmark, but also to make contact with Ref. \onlinecite{BH2Dcorrelationspreading}, we begin with a sudden quench from deep in the Mott insulator phase to the superfluid. In this section we define the energy scale by setting $U=1$. The initial Hamiltonian has zero tunnelling, $J=0$, and the initial ground state is a Fock state:
\be 
\ket{11111111....}
\label{psi_in}
\ee 
with one particle per site. This is a product state that can be represented by an initial iPEPS with bond dimension $1$. Then non-zero tunnelling is suddenly switched on at $t=0$. As in Ref. \onlinecite{BH2Dcorrelationspreading} we consider $1/J=19.6$, i.e., a quench withing the Mott phase. This quench has been performed experimentally in Ref. \onlinecite{exp_sudden_BH_2D} although with a somewhat smoother ramp. 

After the quench we follow time evolution of the single particle correlation function
\be 
C^{sp}_{R} = \frac12 \bra{\psi(t)} b_i^\dag b_j + b_j^\dag b_i \ket{\psi(t)} .
\label{Csp}
\ee 
Here $r$ is a distance between sites $i$ and $j$. Figure \ref{fig:sudden} shows time evolution of the NN correlator, $C^{sp}_{1}$, up to $Jt=0.5$. Acceptable convergence in this time window requires bond dimension at least $D=11...14$. If we were looking just at $C^{sp}_{1}(Jt)$ then it might appear converged already for $D=6...9$ but closer inspection of the corresponding NTU error in the bottom panel of Fig. \ref{fig:sudden} reveals that the NTU error does not improve in this range of $D$ as if adding more bond dimension did not improve expressive power of the iPEPS ansatz for this problem. 
Hidden symmetries may require increasing $D$ not by $1$ but by $2$ or more in order to accommodate not just one more virtual state but a whole multiplet before the expressive power is improved \cite{Hasik_multiplets}.
The error begins to improve again from $D=10$ and already $D=11$ brings it down to an acceptable level. At the same time, the curves $C^{sp}_{1}(Jt)$ appear converging again but this time with an acceptable level of the integrated NTU error. 

This test shows that a combination of the more $D$-efficient NTU algorithm, than the simple update used in Ref. \onlinecite{BH2Dcorrelationspreading}, and higher bond dimensions can significantly increase simulable evolution time. The result encourages us to step beyond the sudden quench and attempt smooth KZ ramps that, by their very nature, take longer times. 

\begin{figure}[t!]
\vspace{-0cm}
\includegraphics[width=0.999\columnwidth,clip=true]{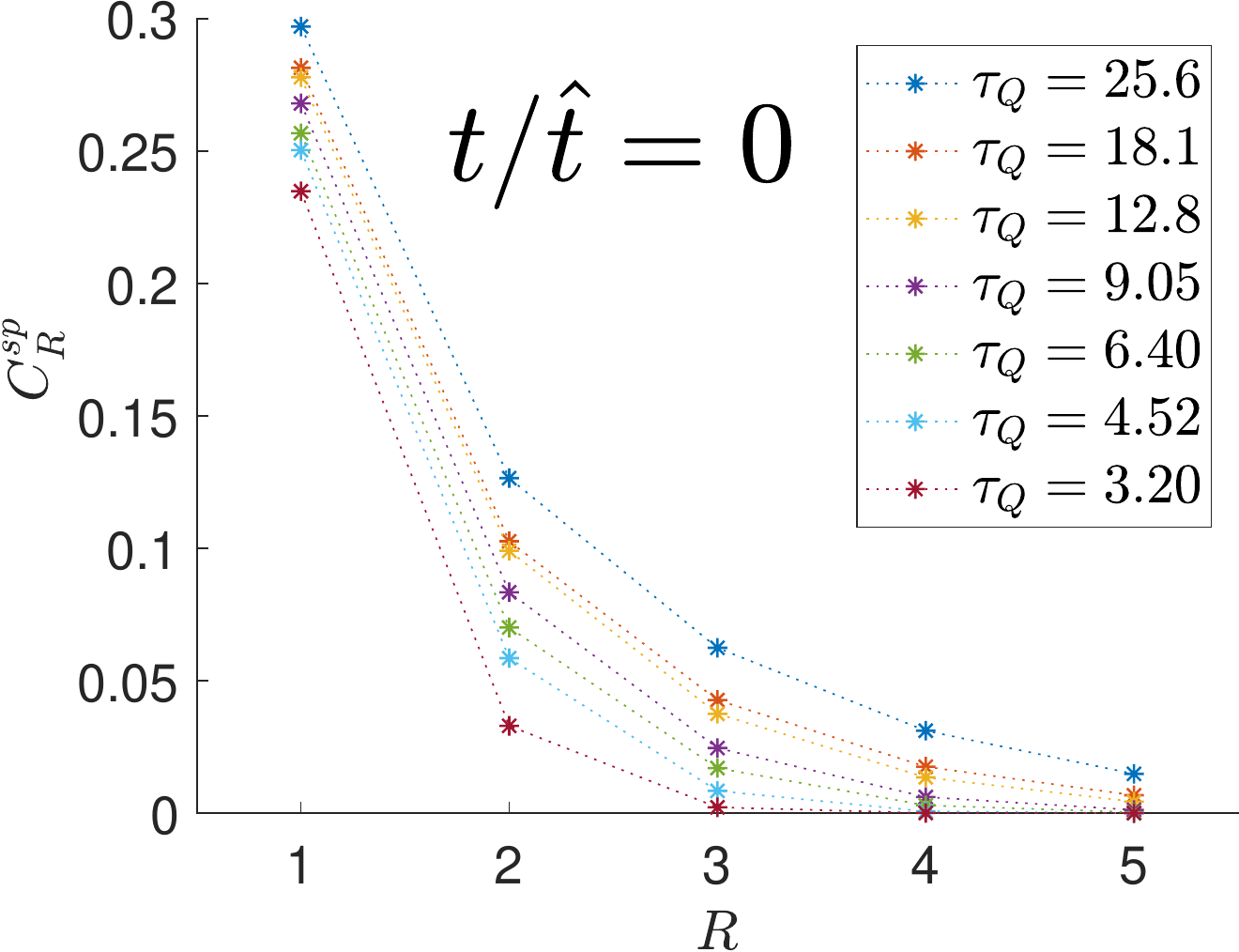}
\includegraphics[width=0.999\columnwidth,clip=true]{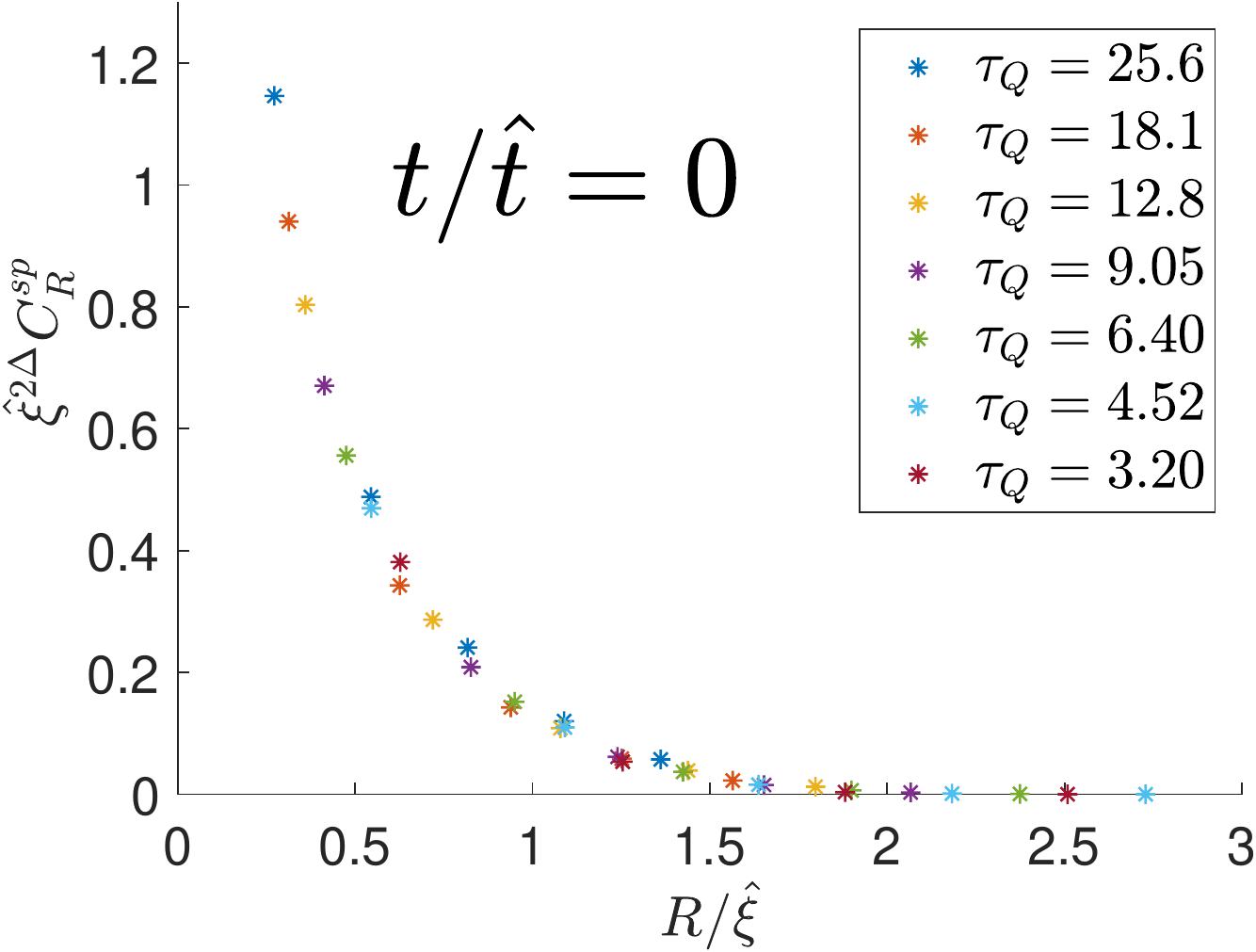}
\vspace{-0cm}
\caption{
{\bf KZ ramp - single particle correlations at ${\bf t=0}$. }
The figure shows the single particle correlation functions at the scaled time $t/\hat t=-1$ for several values of the quench time, $\tau_Q$.
The correlator is scaled according to the more general KZM scaling hypothesis \eqref{eq:KZCr}.
The scaling makes the plots for different $\tau_Q$ collapse to a single scaling function $F_C(-1,R/\hat\xi)$. 
Here we set $U=1$, $Jdt=0.005$, physical dimension $d=3$, and bond dimension $D=14$.
}
\label{fig:Csp_KZ}
\end{figure}

\begin{figure}[t!]
\vspace{-0cm}
\includegraphics[width=0.999\columnwidth,clip=true]{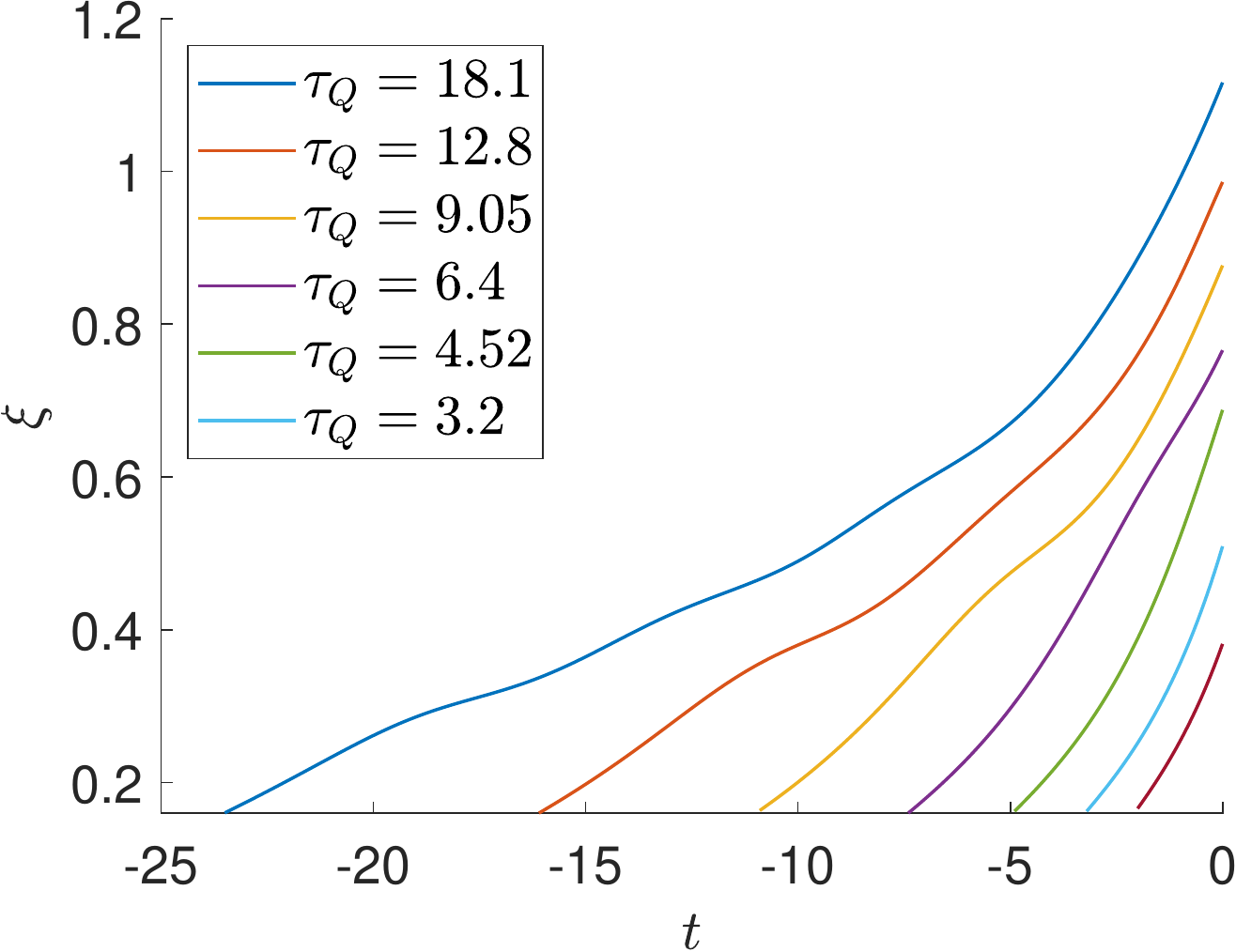}
\includegraphics[width=0.999\columnwidth,clip=true]{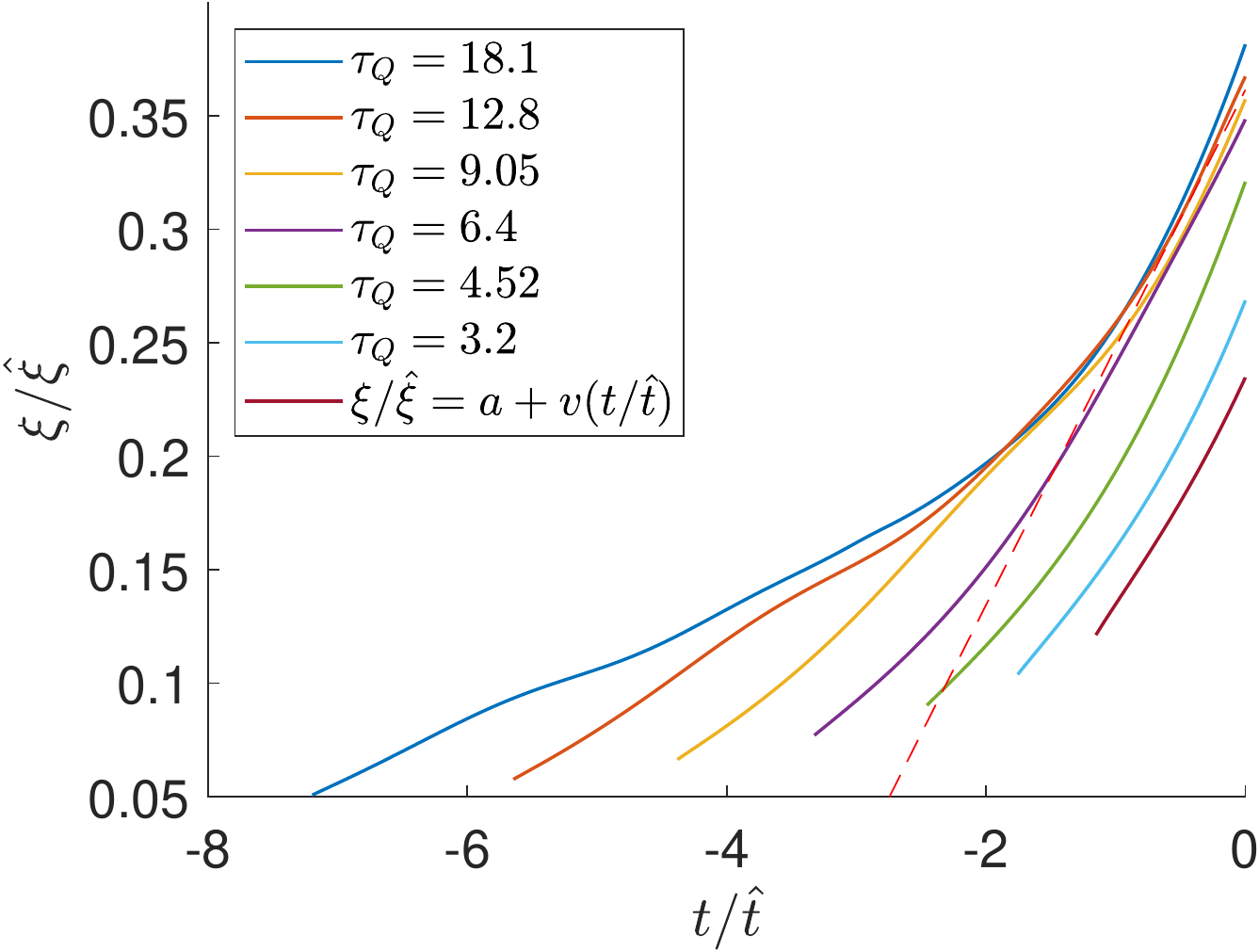}
\vspace{-0cm}
\caption{
{\bf KZ ramp - correlation length. }
The single particle correlation function was fitted with exponents, $C^{sp}_R(t)\approx A\exp(-r/\xi)$, to obtain/define time dependence of the correlation length, $\xi(t)$, during the KZ ramp. 
In (a) bare $\xi(t)$ is shown for several quench times $\tau_Q$.
In (b) scaled correlation length is shown in function of scaled time. 
For the slowest quenches the scaled plots collapse in the KZ regime after $-\hat t$. 
In this regime they are linear fitted with the dashed line.
Its slope yields velocity $v=0.11(3)$ for $(U/J)_c=16.7$ and $U=1$ or,
more generally, $v=1.8(5)J$ for $(U/J)_c=16.7$.
}
\label{fig:xi_KZ}
\end{figure}

\section{Kibble-Zurek ramp}
\label{sec:KZramp}

The Kibble-Zurek quench also begins from the product state \eqref{psi_in} but the hopping rate is increased by a smooth ramp instead of the sudden jump. Near the critical point the ramp can be approximated by a linear slope. It is convenient to parameterize the ramp as
\be 
J=J_c \left[ 1 + \epsilon(t) \right],
\label{eq:J_t}
\ee 
where $J_c$ is the critical point and $\epsilon(t)$ is varied from $-1$ to $\infty$ either as a straight linear ramp $\epsilon(t)=t/\tau_Q$ or, for instance,
\be 
\epsilon(t)=
\left\{
\begin{array}{ll}
\frac{t}{\tau_Q}-\frac{4}{27}\frac{t^3}{\tau_Q^3} & {\rm , when}~~ t<0\\
\frac{t}{\tau_Q} & {\rm , when}~~ t\geq0
\end{array}
\right.
\label{eq:epsilon_t}
\ee 
The former is just linear while the latter can be considered approximately linear in the neighborhood of the critical point at $t=0$, where $\epsilon(t)\approx t/\tau_Q$, provided that quench time $\tau_Q$ is long enough for $\hat t$ in \eqref{hatt} to fall withing the regime of validity of the linearization. The additional qubic term in \eqref{eq:epsilon_t} was added to make its first derivative equal to zero at the beginning of the ramp when $t=-3\tau_Q/2$. This smoothing prevents extra initial excitations that would be created by the abrupt beginning of the linear ramp and might overshadow the KZM excitations created near the critical point. They do not pose a problem for long enough $\tau_Q$ when their energy, proportional to $\tau_Q^{-2}$, becomes negligible compared to the KZM excitation energy that is proportional to $\hat\xi^{-3}\propto\tau_Q^{-1.2}$, but extra bond dimension would be necessary from the very beginning of the tensor network simulation in order to accommodate their extra entanglement. In principle the extra entanglement is not a problem for a quantum simulator/experiment but the relative suppression of the abrupt excitation still requires longer ramp times that are limited by dissipation. In either case there are good reasons to begin the ramp smoothly.

Furthermore, as the on-site repulsion strength, $U$, depends on the recoil energy relatively weakly --- when compared to the hopping rate --- here we conveniently assume that it is constant and choose the unit of energy such that $U=1$. Even if we allowed $U$ to be time-dependent it could be linearized near the critical point and the only effect of the time dependence would be effective multiplication of $\tau_Q$ by a constant factor. This factor would not affect the KZM scaling hypothesis.  

In our simulations the tunneling rate is smoothly ramped up to the critical point at $J_c=1/16.7$ with a time step $dt-0.1$ that is short enough for the second order Suzuki-Trotter scheme to be accurate. 
As our aim is to verify the KZM power laws, quench times are incremented geometrically as $\tau_Q=0.1\cdot 2^{m/2}$, where $m$ is a non-negative integer up to $16$. Longer $\tau_Q$ require larger bond dimensions, up to $D=14$, as they allow for longer KZM correlation length $\hat\xi$ to build up. The accuracy/convergence was monitored with the NTU error as for the sudden quench. We present results obtained with the physical dimension $d=3$. Selective test with $d=4$ show that $d=3$ is accurate enough in consistency with small variance of occupation numbers in our simulations.

Our main focus is the single particle correlation function. It is the most sensitive probe of the KZM as it quantifies just how the long range order builds up when the system is driven across the Mott-superfluid transition. In particular, according to the general KZM scaling hypothesis \eqref{KZscalingO}, when the ramp is crossing the critical point at $t=0$ the correlator should satisfy:
\be 
\hat\xi^{2\Delta_{sp}} C^{sp}_R(t=0) = f_C\left(R/\hat\xi\right).
\label{eq:KZCr_crit}
\ee 
Here $f_{C}$ is a non-universal scaling function, $\Delta$ is an anomalous dimension, and $\hat\xi\propto\tau_Q^{\nu/(1+z\nu)}$ is the KZ correlation length. The correlator at the critical point is plotted in Fig. \ref{fig:Csp_KZ}. The top panel shows raw data for $C^{sp}_R(t=0)$ while the bottom one the same data but scaled according to \eqref{eq:KZCr_crit}. In the rescaling we use $\hat\xi=1\cdot\tau_Q^{\nu/(1+z\nu)}$ and $\hat t=1\cdot \hat\xi^z$ with the numerical coefficients set equal to $1$ for definiteness. For the single particle correlation function $2\Delta_{sp}=1+\eta$, where $\eta=0.038176(44)$~\cite{ElstnerMonien,BHM_Svistunov,Krutitsky_review}. The collapse of the plots with different $\tau_Q$ demonstrates that we reached quench times long enough for the KZM scaling hypothesis to hold as their $\hat t$ is small enough to fall within the critical regime near the transition.

Although the correlation function is not quite exponential, an exponential profile seems to be a reasonably good first approximation that allows to characterize the range of correlations by a single number. In order to ignore numerical noise in the correlator's long range tail we define the correlation length as $\xi(t)=\ln{C^{sp}_1(t)/C^{sp}_2(t)}$. The length is plotted in the top panel of Fig. \ref{fig:xi_KZ} for several different quench times. Furthermore, motivated by a more general KZM scaling hypothesis,
\be 
\hat\xi^{2\Delta} C^{sp}_R(t) = F_C\left(t/\hat t,R/\hat\xi\right),
\label{eq:KZCr}
\ee 
that should hold in the KZM regime after $-\hat t$, in the bottom panel of Fig. \ref{fig:xi_KZ} we show scaled correlation length, $\xi(t)/\hat\xi$, in function of scaled time, $t/\hat t$. According to the hypothesis, for long enough $\tau_Q$ the scaled plots should collapse in the KZM regime and, indeed, this is what we can see for the slowest quenches. The collapse allows a linear fit to the collapsed sections of the plots after $-\hat t$. Our estimate of the slope is $v=1.8(5)J$. According to the causality version of KZM, the slope is upper bounded by twice the sound velocity at the critical point and, indeed, it is lower than the Lieb-Robinson velocity $6(2)J$ predicted and measured in Refs. \onlinecite{BH2Dcorrelationspreading,exp_sudden_BH_2D}, respectively. However, it is strangely low as compared to the upper bound, at odds with many other examples \cite{sonic}. We will come back to this issue below.

\begin{figure}[t!]
\vspace{-0cm}
\includegraphics[width=0.999\columnwidth,clip=true]{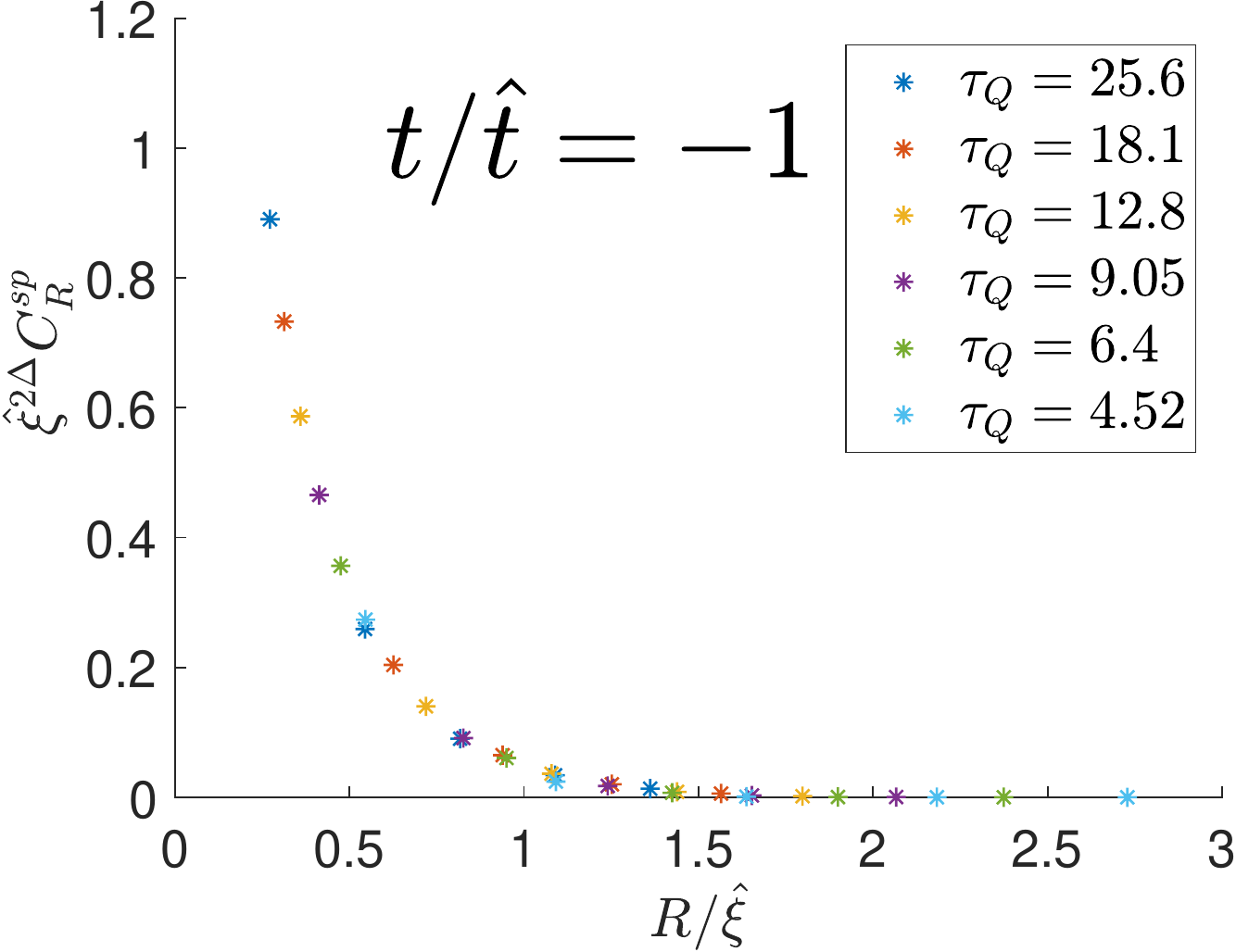}
\vspace{-0cm}
\caption{
{\bf KZ ramp - single particle correlations at ${\bf -\hat t}$. }
The plot shows scaled single particle correlation functions at $-\hat t$, 
where $\hat t=1\tau_Q^{z\nu/(1+z\nu)}$. 
The scaling makes the plots for different quench times collapse to a single scaling function $F_C\left(-1,R/\hat\xi\right)$. 
Here $U=1$, $Jdt=0.005$, physical dimension $d=3$, and bond dimension $D=14$.
}
\label{fig:Csp_KZ_minusthat}
\end{figure}

In the meantime, we observe that the collapse in the bottom panel of Fig. \ref{fig:xi_KZ} is consistent with the general KZM scaling hypothesis \eqref{eq:KZCr}. This conclusion is further corroborated by a direct test --- without any assumption of an exponential or any other specific profile ---  made in Fig. \ref{fig:Csp_KZ_minusthat}, where scaled correlation functions for different $\tau_Q$, but for the same scaled time $t/\hat t=-1$, are plotted together. Their collapse appears even better than the later one at $t/\hat t=0$ in Fig. \ref{fig:Csp_KZ}. These earlier states are less entangled, their correlations are shorter, hence their representation by the tensor network is more accurate. Similar collapses can be obtained in the whole range $t/\hat t\in[-1,0]$ completing demosntration of the KZM scaling hypothesis for the single particle correlation function.

\begin{figure}[t!]
\vspace{-0cm}
\includegraphics[width=0.999\columnwidth,clip=true]{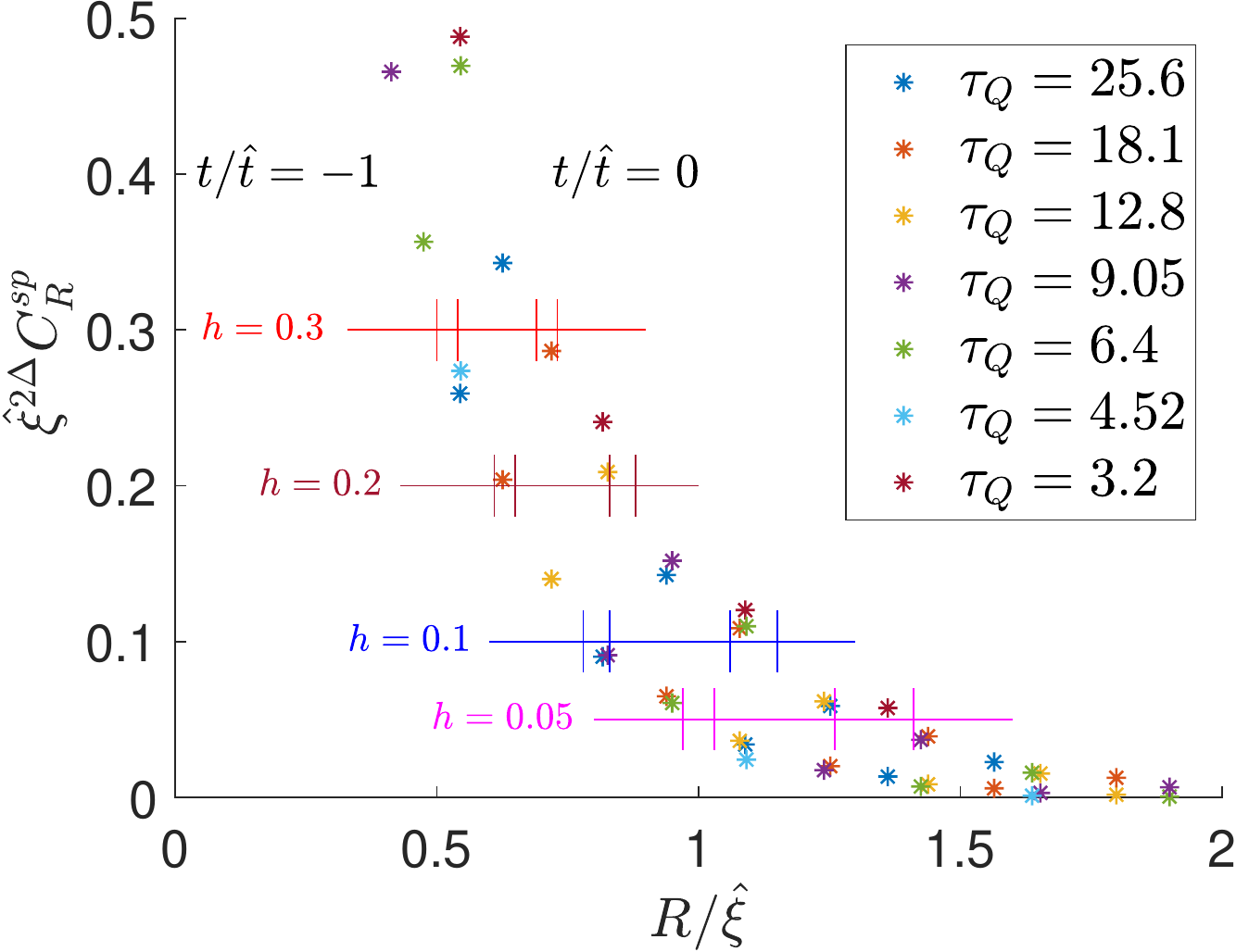}
\vspace{-0cm}
\caption{
{\bf KZ ramp - correlation growth. }
Here we collect together the collapsed scaled correlation functions at the scaled times $t/\hat t=-1,0$ in Figs. \ref{fig:Csp_KZ_minusthat} and \ref{fig:Csp_KZ}.
Horizontal lines mark values of threshold $h$ in Eq. \eqref{eq:h} that are used to estimate the increase of the correlation range between the two scaled times. Each pair of vertical line segments delimits the range of scaled correlation distance, $R/\hat\xi$, where the horizontal line is estimated to cross with the collapsed scaled correlation function at either $t/\hat t=-1$ or $t/\hat t=0$. For each $h$ the difference between the two distances is the speed at which the correlation range is growing between the two scaled times. The speeds are listed in Table \ref{tab:v_h} together with their error bars. 
}
\label{fig:speed}
\end{figure}

The collapsed correlation functions in Figs. \ref{fig:Csp_KZ} and \ref{fig:Csp_KZ_minusthat}, equal to the scaling functions in \eqref{eq:KZCr}, provide a more controlled way to estimate the propagation speed \cite{sonic}. For a small threshold value $h>0$ and the two values of the scaled time, $t/\hat t=-1,0$, equation
\be 
F_C\left(t/\hat t,R/\hat\xi\right) = h
\label{eq:h}
\ee 
can be solved with respect to scaled distance $R/\hat\xi$. Given that for $z=1$ we have $\hat t=\hat\xi$, the increase of the scaled distance between $t/\hat t=-1$ and $t/\hat t=0$ is the propagation speed, $v(h)$. Gradually decreasing $h$ allows to probe the speed at which farther correlations are spreading and thus make contact with the Lieb-Robinson bound on the asymptote of the correlation function. Figure \ref{fig:speed} shows graphic solution of \eqref{eq:h}, including its error bars, that results in a series of estimates: $v(0.3)=0.19(4)$, $v(0.2)=0.23(5)$, $v(0.1)=0.30(7)$, $v(0.05)=0.34(11)$ in our units where $U=1$. The same speed estimates for an arbitrary $U$ are listed in Table \ref{tab:v_h}. The speed appears to increase as threshold $h$ is lowered but, at the same time, its error bars increase due to the growing relative significance of numerical uncertainties farther in the correlator tail. Within the error bars the speed is approaching the estimate $6(2)J$~\cite{BH2Dcorrelationspreading,exp_sudden_BH_2D} that is its upper speed limit according to the casual picture of the Kibble-Zurek mechanism.

 \begin{table}[h!]
     \centering
     \begin{tabular}{|c|c|}
     \hline
       h   & v \\
       \hline
       0.3  & 3.2(7)J \\
       0.2  & 3.8(8)J \\ 
       0.1  & 5.0(11)J \\ 
       0.05 & 5.6(17)J \\    
       \hline
     \end{tabular}
     \caption{ 
     {\bf KZ ramp - correlation growth. } 
     The speed at which the single particle correlations are spreading in the KZ regime estimated in Fig. \ref{fig:speed} for decreasing values of threshold $h$ in Eq. \eqref{eq:h}. The brackets enclose maximal error bars of the last digit. 
     Its upper speed limit is $6(2)J$ according to Refs. \onlinecite{BH2Dcorrelationspreading,exp_sudden_BH_2D}.
     \label{tab:v_h}}
 \end{table}

\begin{figure}[t!]
\vspace{-0cm}
\includegraphics[width=0.999\columnwidth,clip=true]{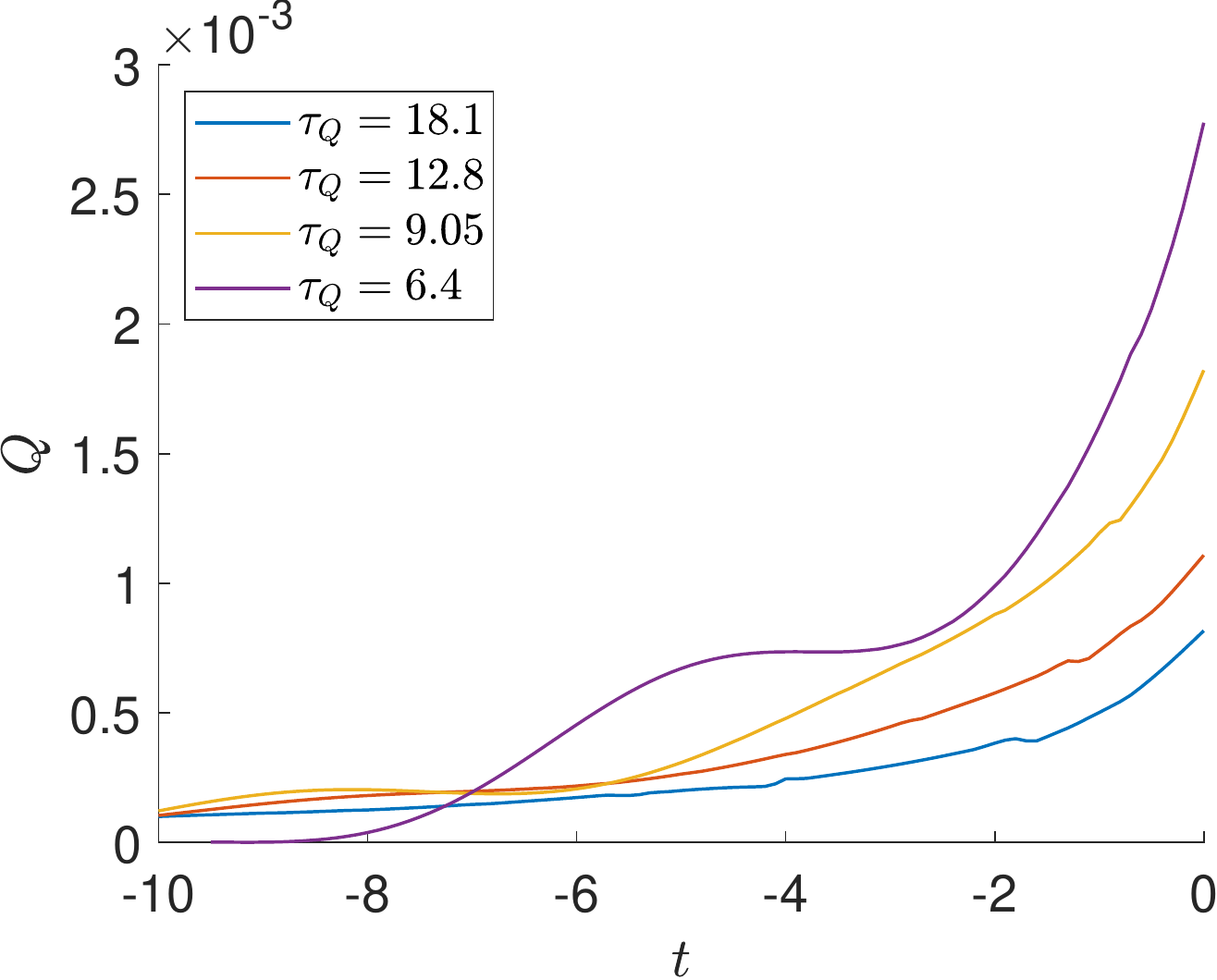}
\includegraphics[width=0.999\columnwidth,clip=true]{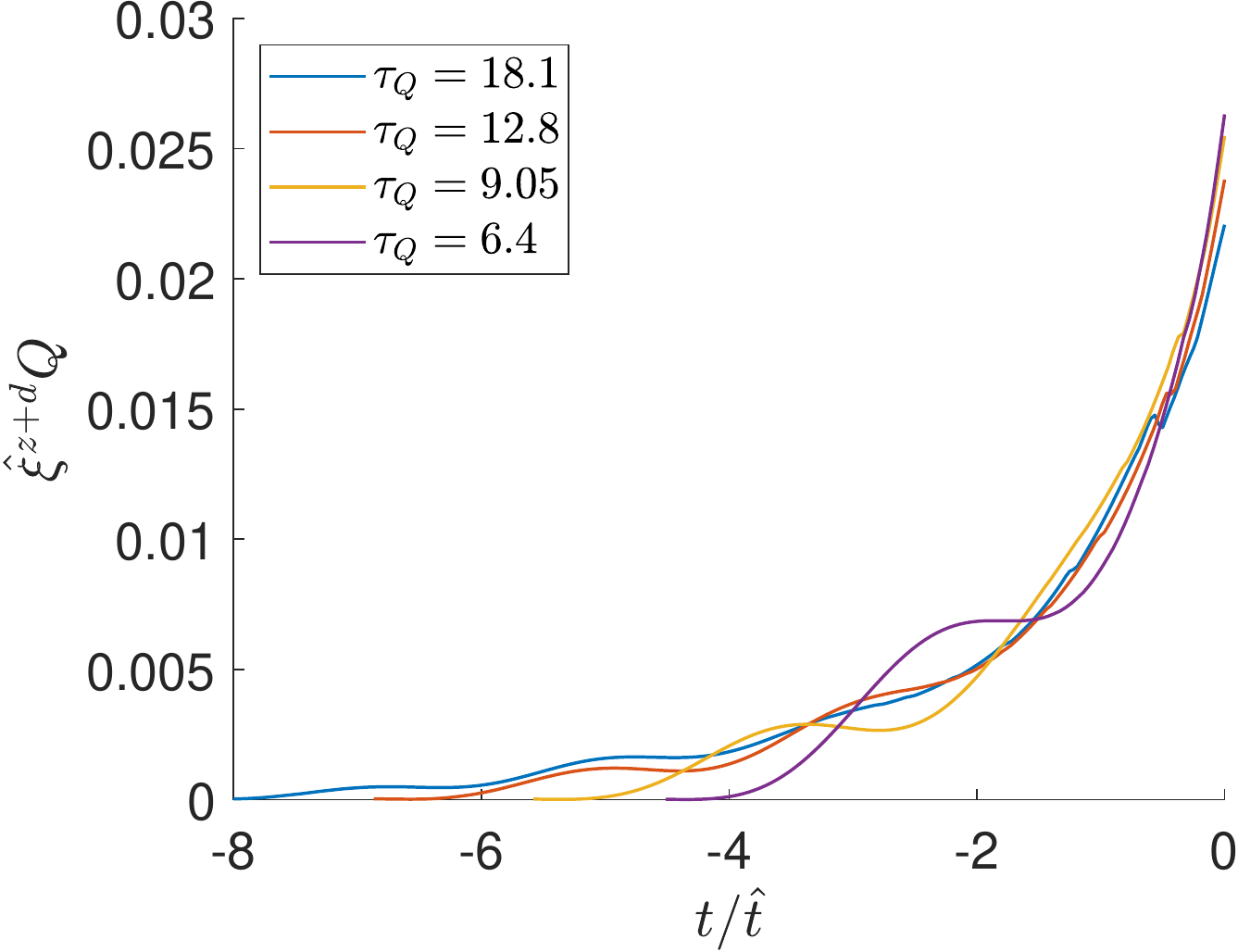}
\vspace{-0cm}
\caption{
{\bf KZ ramp - excitation energy per site. }
Both panels show the excitation energy per site. 
In the top panel bare energy $Q$ is shown in function of time $t$.
In the bottom panel both the energy and the time are scaled according to the KZM scaling hypothesis \eqref{eq:Q_t}.
The scaling make the plots with different $\tau_Q$ collapse in the KZM regime after $t/\hat t=-1$.
Here we set $U=1$, $Jdt=0.005$, physical dimension $d=3$, and bond dimension $D=14$.
}
\label{fig:Q}
\end{figure}

In addition to the single particle correlation function we can also consider excitation energy per site: 
\be 
Q(t) = \lim_{N\to\infty} N^{-1} \left[ \bra{\psi(t)} H(t) \ket{\psi(t)} - E_{\rm GS}(t)\right]. 
\ee 
Here $E_{\rm GS}(t)$ is the ground state energy of the instantaneous Hamiltonian $H(t)$ and $N$ is the number of lattice sites. In the KZ regime after $-\hat t$ the excitation energy should satisfy a scaling hypothesis
\be 
\hat\xi^{z+d} Q(t) = F_Q\left( t/\hat t \right),
\label{eq:Q_t}
\ee 
where $F_Q$ is a non-universal scaling function. On the one hand, with $z+d=3$ the dependence of $Q$ on $\tau_Q$ is very steep allowing for a clear-cut test but, on the other hand, with increasing $\tau_Q$ the excitation energy quickly becomes a small difference of two large numbers that is prone to numerical errors. Nevertheless, in the top panel of Fig.\ref{fig:Q} we plot the excitation energy in function of time for several values of the quench time and in the bottom panel we show the same plots but after the rescaling. The scaled plots demonstrate a rather convincing collapse in the KZM regime after $t/\hat t=-1$.

\section{Thermalization}
\label{sec:TKZ}

In order to follow thermalization in the non-integrable model the KZM ramp can be stopped either on the superfluid side of the transition or even right at the critical point where the thermalization should be the most expedient, unhampered by any gap in the energy spectrum. The following unitary evolution with the critical Hamiltonian conserves the KZM excitation energy density $Q\propto\hat\xi^{-(z+d)}$ while the state evolves into a thermal one with temperature $T$. 
The critical dispersion, $\omega\propto k^z$, means thermal excitations up to $k_T\propto T^{1/z}$ and thermal excitation energy $U_T\propto T^{(z+d)/z}$. Equating $Q$ with $U_T$ we obtain a ``KZ temperature''
\be 
T_{\rm KZ} \propto \hat\xi^{-z} \propto \tau_Q^{-z\nu/(1+z\nu)}
\label{TKZ}
\ee 
and a thermal correlation range $\xi_T\propto k_T^{-1} \propto \hat\xi$. Despite this proportionality the thermal correlator is not the same as the KZ one immediately after stopping the ramp. Interestingly, similar thermalization at and near the critical point but after a sudden quench was considered in Ref. \onlinecite{Arnab_PRX2021}.

\section{Conclusion}
\label{sec:conclusion}

The state of the art quantum simulators of the Bose-Hubbard model at commensurate filling allow one to follow spreading of correlations after a sudden quench for times long enough to estimate their propagation velocities. Our 2D tensor network simulations demonstrate that the experimental times would also be long enough to test the quantum Kibble-Zurek mechanism by verifying the KZM scaling hypothesis for the single particle correlation function. 
The experiment could push this test beyond the limited range of quench times achievable by the classical simulation where the KZM scaling hypothesis should become even more convincing. It could also follow thermalization of the KZ excitations after the ramp is stopped which is a notoriously difficult task for the classical simulation due to rapid growth of entanglement. 
These are the challenges worthy a genuine quantum simulation.

\acknowledgements
%
We are indebted to Ryui Kaneko for comments on the speed limit for correlations.
This research was supported in part by the National Science Centre (NCN), Poland under project 2019/35/B/ST3/01028 (J.M.) and
project 2021/03/Y/ST2/00184 within the QuantERA II Programme that has received funding from the European Union Horizon 2020 research and innovation programme under Grant Agreement No 101017733 (J.D.).
The research was also supported by a grant from the Priority Research Area DigiWorld under the Strategic Programme Excellence Initiative at Jagiellonian University (J.D.).

\bibliography{KZref.bib} 

\end{document}